# Tailoring modal properties of inhibited-coupling guiding fibers by cladding modification

Jonas H. Osório[1], Matthieu Chafer[1, 2], Benoît Debord[1, 2], Fabio Giovanardi[3], Martin Cordier[4], Martin Maurel[1, 2], Frédéric Delahaye[1, 2], Foued Amrani[1, 2], Luca Vincetti[3], Frédéric Gérôme[1, 2], Fetah Benabid[1, 2, *]

[1]*GPPMM Group, XLIM Research Institute, CNRS UMR 7252, University of Limoges, Limoges, France*
[2]*GLOphotonics S.A.S., 1 avenue d'Ester, Ester Technopôle, Limoges, France*
[3]*Department of Engineering "Enzo Ferrari," University of Modena and Reggio Emilia, 41125 Modena, Italy*
[4]*Laboratoire de Traitement et Communication de l'Information, Télécom ParisTech, Université Paris-Saclay, 75013 Paris, France*
*\*Corresponding author: f.benabid@xlim.fr*

**Abstract:** Understanding cladding properties is crucial for designing microstructured optical fibers. This is particularly acute for Inhibited-Coupling guiding fibers because of the reliance of their core guidance on the core and cladding mode-field overlap integral. Consequently, careful planning of the fiber cladding parameters allows obtaining fibers with optimized characteristics such as low loss and broad transmission bandwidth. In this manuscript, we report on how one can tailor the modal properties of hollow-core photonic crystal fibers by adequately modifying the fiber cladding. We show that the alteration of the position of the unity-tubes in the cladding of tubular fibers can alter the loss hierarchy of the modes in these fibers, and exhibit salient polarization propriety. In this context, we present two fibers with different cladding structures which favor propagation of higher order core modes – namely $LP_{11}$ and $LP_{21}$ modes. Additionally, we provide discussions on mode transformations in these fibers and show that one can obtain uncommon intensity and polarization profiles at the fiber output. This allows the fiber to act as a mode intensity and polarization shaper. We envisage this novel concept can be useful for a variety of applications such as hollow core fiber based atom optics, atom-surface physics, sensing and nonlinear optics.

## 1. INTRODUCTION

Intense efforts have been devoted to hollow-core photonic crystal fiber (HCPCF) research since its first proposal in 1995 [1]. In the latest years, HCPCFs have revealed themselves as a great platform for the understanding of the waveguiding mechanisms and as an excellent tool for addressing diverse applications needs.

HCPCF can guide light by photonic bandgap (PBG) [1] or Inhibited-Coupling (IC) [2] mechanisms. In PBG guiding fibers, and similarly to total-internal reflection fibers, the coupling of the core mode to the cladding is forbidden because the cladding modal spectrum is void from any propagation mode at the core guided-mode effective index-frequency space [1]. Otherwise, in IC guiding fibers, the core and cladding modes coupling is robustly minimized by having a strong mismatch in their transverse spatial phases and a small spatial overlap between their fields [2]. In this context, the confinement loss (CL) in IC fibers, in contrast with PBG ones, is strongly dependent on the core contour characteristics. This observation motivated the introduction of the hypocycloid-core contour (*i.e.* negative curvature) in 2010 [3, 4], which allowed to attain a dramatic reduction in the transmission losses in IC guiding HCPCFs.

Thus, a great interest was observed on the development of HCPCFs with hypocycloid-shaped cores and, in particular, on the study of single-ring tubular lattice (SR-TL) HCPCF [5]. The growing interest in this sort of fibers has been motivated by their noteworthy properties, which encompass a cladding geometric simplicity combined with the absence of connecting nodes, and allows obtaining, by virtue of IC criteria, ultralow-loss and broad spectral transmission bandwidth [6]. In this framework, research papers have been published on the optimization of SR-TL HCPCF design to reduce confinement and bend losses [6-8], to study their modal content and to achieve single-mode operation [6, 9, 10]. Additionally, one can find in the literature a set of investigations on the use of such fibers in applied fields as in mid-IR lasers [11], generation of single-cycle pulses [12], and sensing [13].

SR-TL HCPCFs transmit light by IC guiding [2]. In SR-TL HCPCFs, the lattice tubes define an hypocycloid core contour, which lowers the spatial overlap between core and cladding modes. Additionally, it is endowed with a silica microstructure which is void of structural nodes (which supports low azimuthal number modes and disfavors IC guidance) [6].

As in SR-TL HCPCFs the core contour is demarcated by the tubes that forms the fiber cladding, the definition of their geometric parameters are of great importance for designing the fiber properties. For instance, by adequately choosing the diameter, the thickness and the number of tubes in the cladding, the researcher can design fibers with different core sizes, predict the spectral location of the transmission bands and the losses levels which are suitable for the desired application [6, 14].

Here, in a different fashion, we study and demonstrate that the alteration of the azimuthal position of the cladding tubes can favor the propagation of higher order core modes, *i.e.*, by adequately choosing the spacing of selected cladding tubes, it is possible to tailor the fiber modal properties and alter the loss hierarchy of the modes in the fiber. To achieve this goal, we used the findings of the detailed study on the Poynting vector in the transverse plane of SR-TL HCPCF available in [6], which concluded that the power leakage through the spacing between the lattice tubes is strongly increased when this spacing between them is enlarged.

Thus, here, we explore this concept both theoretically and experimentally by studying and developing two tubular fibers whose lowest loss modes are the $LP_{11}$ and $LP_{21}$, instead of the $LP_{01}$. We believe that such control feature on the mode loss hierarchy is unprecedented in optical fiber.

Additionally, we present experiments on mode transformations in these fibers with modified cladding structures. We show that interesting output intensity profiles can be obtained. In particular, for one of the studied fibers, the superposition of $LP_{01}$ and $LP_{11}$ modes entails an $LP_{11}$-like intensity profile in the fiber output with unusual orthogonal polarization sites. Remarkably, if the input light polarization is conveniently adjusted, these orthogonal polarization regions can be individually or simultaneously excited. We envisage that these properties can be useful for atom optics and sensing experiments.

## 2. SIMULATION RESULTS

Fig. 1 summarizes numerical simulation results for three SR-TL HCPCFs fiber designs (FD). The first FD (see top left of the panel Fiber Design #0 of Fig. 1a) is a SR-TL HCPCF with 10 tubes and constant spacing between the tubes, used here as a reference to study the properties of the novel designs proposed in the following. The second FD consists of a cladding with 8 tubes and two larger gaps defining an angle of 180º between them (Fiber Design #1) and a third fiber with 8 tubes and four larger gaps apart 90º with respect to each other (Fiber Design #2). The fibers have same core diameter (45 µm) and same tubes sizes (outer diameter D = 15 µm and thickness t = 750 nm). FD #1 is obtained from FD #0 structure by simply removing two tubes on the horizontal axis. The obtained larger gaps measure 20.2 µm. FD #2, in turn, has the same number of tubes than FD #1 but a different azimuthal distribution of the same. Thus, in FD #2, we obtain two pairs of larger gaps, one along the horizontal direction and the other along the vertical one. The larger gaps in FD #2 are narrower than in FD #1 and they measure 12.3 µm.

In Fig.1 each of the FD panels shows the fiber structure transverse profile (top left), its effective index (top right), the CL spectrum of the most representative core modes that can be guided through the fibers – namely the $LP_{01}$, $LP_{11}$ and $LP_{21}$ modes (medium of the panel). The intensity profiles of these modes are shown in bottom left. The bottom right of the panel shows the corresponding Poynting vector transverse component. It is worth saying that the LP modes are not true modes of the waveguide [16]. Alternatively, they arise from linear superpositions of the $TE_{01}$, $TM_{01}$ and $HE_{21}$ vectorial modes, which have donut-shaped intensity profiles and not linear polarization [16, 17]. Indeed, the excitation of vectorial modes are more difficult to be experimentally accomplished and the linearly polarized modes are more efficiently excited by usual gaussian beams [17]. Thus, as here we detected superpositions of the linearly polarized modes, we maintain the LP notation for simplicity. Since LP modes with same azimuthal number can exhibit different field distributions, here we define as $LP_{11a}$ the mode with the two lobes along the vertical direction and as $LP_{11b}$ the mode having the two lobes along the horizontal one. Moreover, it is true that for each field distribution there are two different polarizations. However, since the numerical results show a weak dependence on the polarization in the investigated fibers, we only consider the vertically polarized LP modes in our analyses for simplicity.

The effective indices plots show comparable dispersion curves for the three FD. In contrast, the plots for the CL reveal that the alteration of the azimuthal position of the cladding tubes in the fiber structure allows changing the modes losses hierarchy. For FD #0, as expected, the simulation results show that the $LP_{01}$ is the core mode with lowest CL figures (with losses around 3.6 dB/km at 1000 nm), followed by the higher order modes. Instead, we see that, for FD #1, the inclusion of larger gaps between the tubes at 180º enhances more than two decades the CL for the fundamental mode $LP_{01}$ and for the modes $LP_{11b}$, while keeping low impact in the CL of the $LP_{11a}$ and $LP_{21}$ modes. It entails that, for FD #1, the mode with lowest loss is the $LP_{11a}$ mode, whose CL is calculated to be 33 dB/km at 1000 nm – not so different than the loss of the same mode in FD #1 (12 dB/km).

In this context, it is also remarkable to observe that the fiber asymmetry in FD #1 entails very different loss figures among the modes of $LP_{11}$ family. It is noteworthy that the $LP_{11a}$ mode, which has the zero electric field line separating the two lobes passing through the larger gaps, exhibits much lower losses than the $LP_{11b}$, which has the zero line passing through a narrower gap. The larger gaps between the cladding tubes in the horizontal direction causes the $LP_{11b}$ mode field to spread towards the silica jacket, which strongly deteriorates its confinement and highly increases its loss from 11 dB/km to 16500 dB/km at 1000 nm. Also, the $LP_{21}$ mode has the zero line passing through the larger gaps and, thus, shows a weak dependence on them. At 1000 nm, its CL passes from 70 dB/km in FD #0 to 360 dB/km in FD #1. Finally, it is worthy observing in Fig. 1 that the asymmetry in the FD #1 cladding caused by the larger gaps, enhances the difference between the effective refractive indexes of $LP_{11a}$ and $LP_{11b}$ modes. At 1000 nm, the difference between their effective refractive indices is $7 \times 10^{-5}$.

Furthermore, by observing in Fig. 1 the FD #2 figures, we can observe that the addition of four larger gaps defining an angle of 90º between each other again allows to change the modes losses hierarchy. For this fiber, the inclusion of the bigger gaps at 90º deteriorates the losses of the $LP_{11}$ mode, causing the modes with lowest losses to be the $LP_{21}$ and $LP_{01}$ modes. In particular, at 1000 nm, the mode with the lowest loss is the $LP_{21}$ one with CL around 80 dB/km.

The explanation of how the alteration of the azimuthal position of the cladding unity-tubes can modify the modes losses hierarchy is centered on the fact that the power leakage through the spacing between the lattice unity-tubes is strongly increased when this spacing is larger [6]. To address this point, we calculated the power flux along the fiber radial direction for the most representative modes guided through the fibers. The density of power flowing along the radial direction is accounted by the normalized radial component of the Poynting vector, $p_r$, given by Eq. (1), where $\vec{E}$ and $\vec{H}$ are the electric and magnetic fields of the mode, $\hat{r}$ is the radial unit vector, and $p_z$ the maximum value of the longitudinal component of the Poynting vector [6]. The results of these simulations are shown in Fig. 1, where $p_r$ (at the wavelength of 1000 nm) is plotted in logarithm scale.

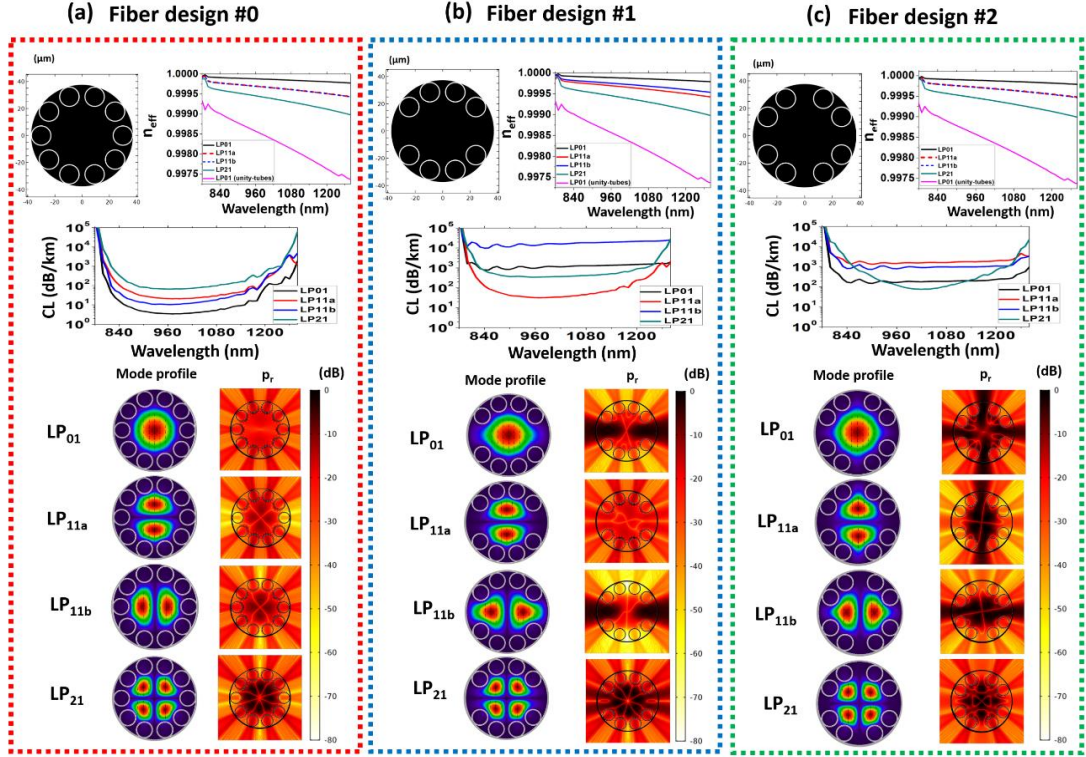

Fig. 1. Schematic diagram for the cross sections of the studied fibers; plots for the effective refractive index ($n_{eff}$) and CL as a function of the wavelength for $LP_{01}$ (black curve), $LP_{11a}$ (red curve), $LP_{11b}$ (blue curve) and $LP_{21}$ (green curve); mode profiles and color map for the radial component of the Poynting vector ($p_r$ – in logarithmic scale) for $LP_{01}$, $LP_{11a}$, $LP_{11b}$ and $LP_{21}$. (a) Fiber design #0: tubular fiber with identical gap between the lattice tubes. (b) Fiber design #1: tubular fiber with two bigger gaps at 180°; (c) Fiber design #2: tubular fiber with four bigger gaps at 90°. In the effective refractive index plots, the unity-tubes $HE_{11}$ ($LP_{01}$) dispersion (pink curves) is shown for comparison.

$$p_r = \frac{1}{2p_z}\vec{E} \times \vec{H^*} \cdot \hat{r} \qquad (1)$$

It is seen that, for FD #0, the main channel for mode leakage is the direction along the lattice tubes instead of the gaps between the same. For the mode $LP_{01}$, the electric field distribution does not depend on azimuthal angle. This results in a symmetric radial power flux distribution on the radial direction. Conversely, the modes with non-zero azimuthal number exhibit one or more zero lines in the electric field distribution. For these modes, the $p_r$ distributions show that the flux along these zero electric field lines is significantly lower. For the mode $LP_{11a}$, this corresponds to the horizontal axis and, for $LP_{11b}$, to the vertical direction. For the mode $LP_{21}$, zero electric field lines exist along horizontal and vertical directions.

In FD #1, the larger spacing between the lattice tubes in the horizontal direction causes the gap between the tubes to be the main channel for power leakage of $LP_{01}$ and $LP_{11b}$. In fact, by comparing CL figures of FD #0 and FD #1 we can observe that the CL of $LP_{01}$ and $LP_{11b}$ significantly increase whereas the variation for the CL of $LP_{11a}$ and $LP_{21}$ is much weaker. As a consequence of that, in FD #1, the $LP_{11a}$ mode becomes the lowest loss one. This property is confirmed by the results of FD #2. In this case, the only mode having zero electric field lines along the direction of the two pairs of larger gaps (and, thus, low $p_r$ along these directions) is the $LP_{21}$ mode. Therefore, it exhibits the lowest CL in this fiber (around 1000 nm wavelength).

To stress that the change in the modes loss hierarchy in these fibers is due to the modification in the fiber cladding rather than any resonant coupling of the core modes to the cladding, we provide in Fig. 1 the dispersion curve for the fundamental mode of the unity-tubes (pink line). We see there is a great difference between the core modes effective refractive indexes and the effective refractive index of the fundamental mode in the unity-tube, which excludes the possibility of resonant coupling between them.

Ergo, we can observe that the CL hierarchy of the fiber modes can be effectively tailored by working on the unity-tubes distribution in the cladding of tubular fibers. In this approach, we take into account the field distributions of the core modes and strategically enlarge selected gaps between the unity-tubes in order to favor the propagation of higher order modes.

## 3. FIBER FABRICATION AND LOSS MEASUREMENT

To experimentally study the results attained in the simulations, we performed the fabrication of two different fibers, which reproduced the characteristics of the ones studied in the last section (Fig. 2a and Fig 3a show their cross sections). The fibers were fabricated by using stack-and-draw technique. Here, we refer to the fabricated fibers as F#1 and F#2 to avoid confusion with the simulated ones. F#1 structure has larger gaps between the tubes defining an angle of 180º between them. F#2 has, instead, larger gaps at 90º between each other. The larger gaps between the tubes in F#1 measure (17.0 ± 0.2) μm while the smaller ones measure (4 ± 1) μm. In F#2, the larger gaps between the lattice tubes measure (9.2 ± 0.8) μm and the smaller ones (3.4 ± 0.4) μm.

Cutback measurements were performed in order to account for $LP_{11}$ and $LP_{21}$ losses in F#1 and F#2 respectively. To achieve this, we optimized the input coupling conditions to obtain the $LP_{11}$ and $LP_{21}$ profiles at the fibers output before performing the cutback. Fig. 2b show the transmitted spectra through F#1 for different fiber lengths. For each fiber length, the output beam profile is recorded to ensure that only the mode of interest is excited. Fig. 2c shows the reconstructed near field profiles for the spectra presented in Fig. 2b. Here, the camera images were superposed to the fiber cross section to help visualization. The results shown in Fig. 2c readily demonstrate that the measured loss (Fig. 2d, blue line) stands for the $LP_{11}$ mode loss in F#1, measured to be 200 dB/km for wavelengths around 1000 nm. Also, it is shown in Fig. 2d the CL for $LP_{11}$ which was simulated from a model based of the fiber microscopy image (red line). A good agreement is seen between simulated and experimental results.

Analogously, Fig. 3b show the transmitted spectra through F#2 for different fiber lengths when the fiber output was that of the $LP_{21}$ mode (near field profiles available in Fig. 3c, with camera images superposed to the fiber cross section to help visualization). The measured loss for the $LP_{21}$ mode in F#2 was accounted as 300 dB/km for wavelengths around 1000 nm (Fig. 3d). Similarly to F#1, the simulated CL for $LP_{21}$ mode in F#2 (red line) compares well with the measured results. It is noteworthy that it is the first time that exciting selectively $LP_{11}$ or $LP_{21}$ over such long section of fiber of this type is reported.

## 4. MODE TRANSFORMATIONS IN MODIFIED CLADDING FIBERS

Although the tubular fibers with modified cladding we study herein can favor the propagation of higher order modes and change the modes losses hierarchy in the fibers, we can, by adequately tuning the coupling conditions, excite combinations of the modes supported by the fiber structure. Indeed, we could, by conveniently tuning the light launching conditions, obtain $LP_{01}$-like and $LP_{11}$-like mode profiles in F#1 output, and $LP_{01}$-like, $LP_{11}$-like, and $LP_{21}$-like mode profiles in F#2 output. The observation of these mode profiles is consistent to the modal content of F#1 and F#2 measured using spectral and spatial imaging technique ($S^2$ technique) [17], which detected that $LP_{01}$ and $LP_{11}$ contributions in F#1 output and $LP_{01}$, $LP_{11}$ and $LP_{21}$ contributions in F#2 output.

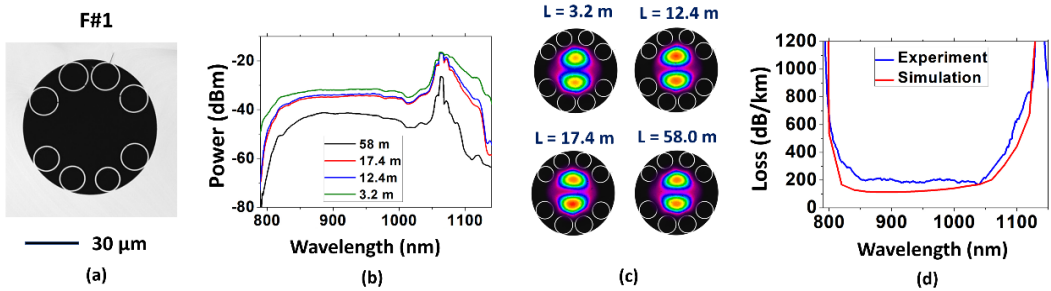

Fig. 2. (a) Cross section of the fiber with two bigger gaps at 180º (F#1), (b) the transmitted spectra and (c) the near field profiles for different lengths. (d) Measured loss for the $LP_{11}$ mode (blue line) together with the simulated CL (red line).

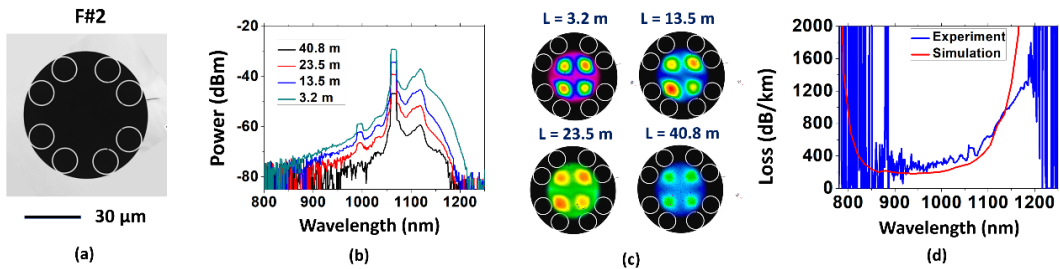

Fig. 3. (a) Cross section of the fiber with four bigger gaps at 90º (F#2), (b) the transmitted spectra and (c) the near field profiles for different lengths. (d) Measured loss for the $LP_{21}$ mode (blue line) together with the simulated CL (red line).

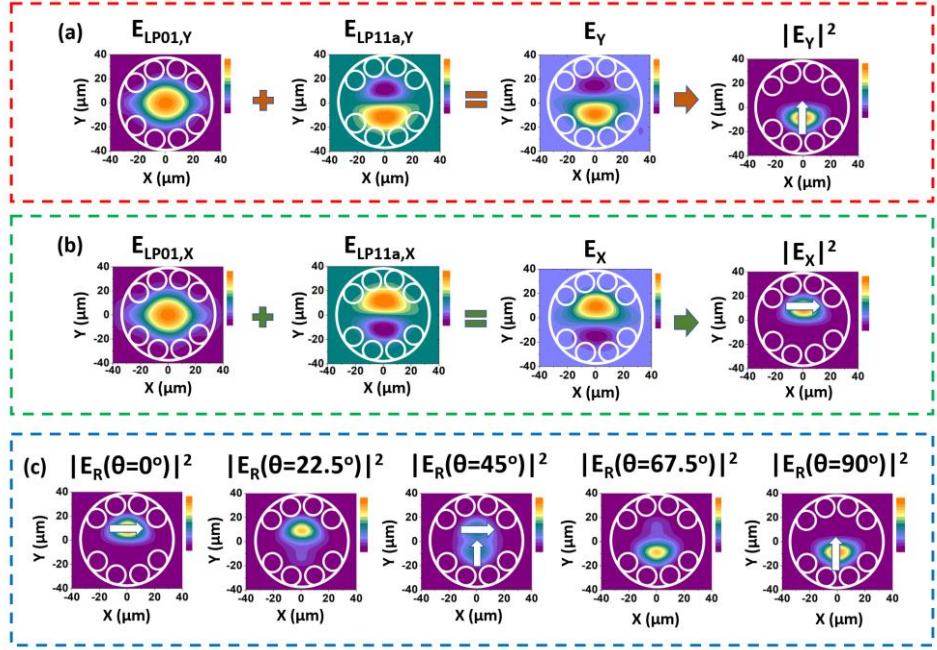

Fig. 4. (a) Electric field profiles of $LP_{01,Y}$, $LP_{11a,Y}$, and of the superposition of these modes in FD #1 ($E_{LP01,Y}$, $E_{LP11a,Y}$ and $E_Y$). $|E_Y|^2$ stands for the intensity profile. (b) Electric field profiles of $LP_{01,X}$, $LP_{11a,X}$, and of the superposition of these modes in FD #1 ($E_{LP01,X}$, $E_{LP11a,X}$ and $E_X$). $|E_X|^2$ stands for the intensity profile. (c) Resultant intensity profile for the superposition of the $LP_{01}$ and $LP_{11a}$ modes as a function of the input polarization angle ($\theta$). Arrows represent the electric field.

In this section, we show that owing to the possibility to excite combinations of these modes on one hand and, on the other, due to their particular polarization properties, we can generate interesting output intensity profiles with unusual intensity profile and polarization distributions. The fiber, thus, acts as a mode intensity and polarization shaper.

As an example, we investigated the superposition of $LP_{01}$ and $LP_{11a}$ modes in FD #1 to illustrate the fiber intensity and polarization shaping capability. Fig. 4a and Fig. 4b show the simulated electric field distributions of vertically and horizontally polarized $LP_{01}$ and $LP_{11a}$ modes ($E_{LP01,Y}$, $E_{LP01,X}$, $E_{LP11a,Y}$ and $E_{LP11a,X}$), and the electric field and intensity profile which results from their superposition ($E_Y$, $E_X$, $|E_Y|^2$ and $|E_X|^2$), calculated according to Eq. (2) and Eq. (3).

$$E_Y = E_{LP01,Y} + E_{LP11a,Y} \quad (2)$$

$$E_X = E_{LP01,X} + E_{LP11a,X} \quad (3)$$

In Fig. 4a, we observe that a y-polarized lobe at the bottom half of the core can be obtained if vertically polarized light is launched into the fiber. Otherwise, Fig. 4b shows that, if horizontally polarized light is coupled into the fiber, an x-polarized lobe at the upper half of the core is obtained.

An interesting outcome of the above properties is seen when we examine a beam with such modal content after passing through a polarizer which axes are defined by an angle $\theta$ with respect to the horizontal plane. The resulted intensity is given by Eq. (4):

$$|E_R(\theta)|^2 = \left[(E_{LP01,X} + E_{LP11a,X})\cos\theta\right]^2 + \left[(E_{LP01,Y} + E_{LP11a,Y})\sin\theta\right]^2 \quad (4)$$

Fig. 4c shows the intensity profiles for different $\theta$ values. It is observed that, as the polarization angle is changed, the output intensity profile is altered so that a $LP_{11}$-like mode profile with unusual orthogonal polarization regions can be obtained when $\theta = 45°$. This is remarkable and very different from the usual electric field configuration of a $LP_{11}$ mode output, whose electric field orientation at the lobes are parallel. When $\theta = 0°$ and $\theta = 90°$, we obtain, respectively, the x- and y-polarized lobes at the upper and bottom halves of the core, which is consistent to Fig. 4a and Fig. 4b results.

The experimental demonstration of this intensity spatial-mode shaping principle was carried for both F#1 and F#2, and is shown in Fig. 5. The setup employed (Fig. 5a) uses a diode laser at 1070 nm to couple into the fiber under test, CCD cameras (C1, C2 and C3) for beam profile recording, and polarizing optical components to control the beam polarization at fiber input and at the different output ports.

Fig. 5b presents the output $LP_{11}$-like intensity profiles measured in C1 for different input polarization angles ($\theta$) for a 3.6 m long F#1 (the camera images were superposed to the fiber cross section to help visualization). We observe that, consistently with the simulations shown in Fig. 4c, the alteration of the input light polarization allows obtaining the bottom lobe, the upper lobe or both lobes in the fiber output. Additionally, Fig. 5c exhibits that, for the situation in which the fiber output has the two-lobed $LP_{11}$-like profile (observed in C1), the detected profiles in C2 and C3 are that of the upper and bottom lobes respectively. As C2 and C3 are placed at different ports of the PBS, we conclude that the lobes in the output profile of the F#1 have orthogonal polarization states, as the simulations have predicted. For comparison, we show in Fig. 5d the profiles detected in C1, C2 and C3 when F#1 is replaced by a 1 m long standard telecom fiber (singlemode at 1550 nm, but which supports few modes at 1070 nm). As expected, the lobes of the $LP_{11}$ mode in the telecom fiber have parallel polarization direction and, therefore, the mode profile is not split by the PBS. The property of having an $LP_{11}$-like output profile with orthogonal polarization sites is, therefore, remarkable and very particular to the tubular fiber design we study herein.

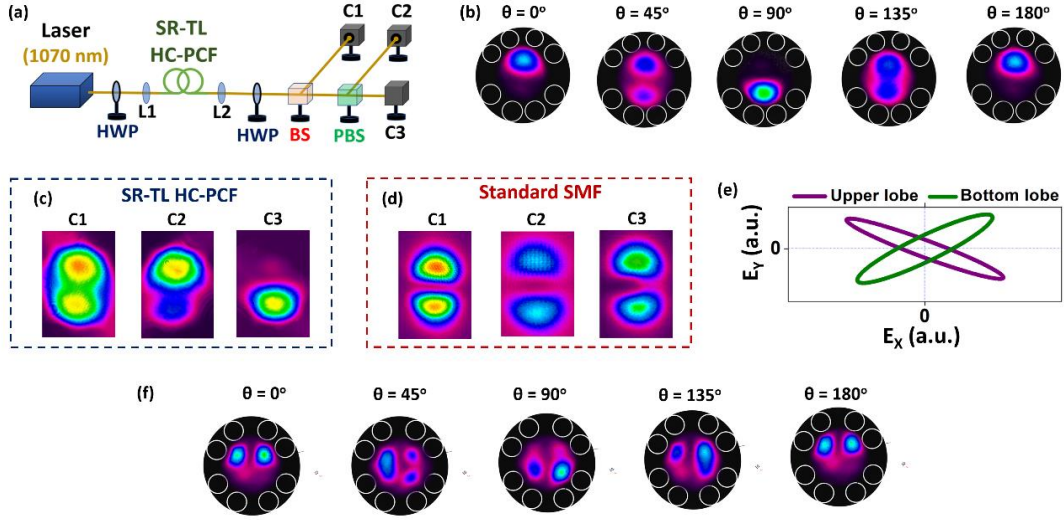

Fig. 5. (a) Experimental setup for the mode transformations characterizations. HWP: half-wave plate; L1 and L2: lenses; BS: beam splitter; PBS: polarizing beam splitter; C1, C2 and C3: CCD cameras. (b) Output intensity profile of F#1 measured in C1 for different input polarization angles. Output intensity profiles measured in C1, C2 and C3 for (c) F#1 and for (d) a standard telecom optical fiber. (e) Polarization ellipses from polarimeter data for upper and bottom lobes in the $LP_{11}$-like profile of F#1 output. (f) Output intensity profile of F#2 measured in C1 for different input polarization angles. In (b) and (f) the camera images were superposed to the fiber cross section for better visualization.

Additionally, we measured, by replacing C1 by a polarimeter, the polarization state of each lobe in the asymmetric SR-TL HCPCF output profile individually. It was done by selectively blocking the $LP_{11}$-like profile lobes and launching the resulting light into the polarimeter. Fig. 5e shows the polarization ellipses measured in the polarimeter for both lobes. The ellipticity ($\varepsilon$) of the polarization ellipses was measured as 0.15 and 0.21 for the upper and bottom lobe respectively. The crossed orientation of the polarization ellipses shown in Fig. 5e confirms that the lobes in F#1 output intensity profile have indeed mutually orthogonal polarization states.

Furthermore, we show in Fig. 5f that it is also possible to tune the output intensity profile shape when using F#2. For this one, we see that the rotation of the input polarization angle ($\theta$) allows to obtain two lobes at the bottom or at the upper part of the core at the fiber output, which, once again, is very distinctive. By simply rotating the polarization angle, we can obtain a variety of intensity profile such as ones with two and three lobes.

## 5. CONCLUSION

In conclusion, we theoretically and experimentally investigated the properties of IC guiding tubular fibers with modified cladding structures. A strategic modification of the azimuthal position of the unity-tubes in the cladding of tubular fiber allows altering the mode losses hierarchy in these fibers. To study this new concept, we studied two novel fibers with modified cladding structures, which favored the propagation of $LP_{11}$ and $LP_{21}$ modes. In addition, we showed that it is possible to work on the combinations of the modes in the fibers with modified cladding in order to obtain interesting intensity and polarization profiles at the fiber output. We believe that this concept will be useful for new experiments in HCPCF based sensing, atom optics, atom-surface interaction and nonlinear optics. For example, in alkali atom filled HCPCF experiments such as the ones reported in [1], the ability to excite the atoms with such intensity profile could be useful to assess the contribution of the atom-surface interaction compared to a Gaussian-like profile. Also, the capability to excite selectively such polarization dependent modes would open the control parameter space in nonlinear optical experiments. Furthermore, the results reported herein reinforce the idea that a deep understanding of the fiber cladding is key for designing IC guiding optical fibers with the desired characteristics and performances.

## ACKNOWLEDGEMENTS

This work has the financial support from BPI and Region Nouvelle Aquitaine under the project 4F, and from ERC H2020 under the project HIPERDIAS Grant Agreement No 687880.

## AUTHORS CONTRIBUTIONS

JHO, MCh, BD, FD, FA, and FGe worked on the fiber fabrication. JHO, FGi, and LV worked on the simulations. JHO, MM, and FD performed $S^2$ measurements. JHO and MCo worked on the model for the modal transformations. JHO performed the experiments on the modal transformations and wrote the paper. FB coordinated the research project. All the authors discussed the results and reviewed the manuscript.

## ADDITIONAL INFORMATION

**Competing financial interests**: The authors declare no competing financial interests.

**Data availability:** The data that support the findings of this study are available from the corresponding author upon reasonable request.